\documentclass[preprint]{aastex}

\usepackage{longtable}
\usepackage{natbib}

\shorttitle{Unification scheme of radio galaxies and quasars falsified}
\shortauthors{Singal and Singh}

\begin{document}

\title{Unification scheme of radio galaxies and quasars falsified by their observed size distributions}

\author{Ashok K. Singal and Raj Laxmi Singh}
\affil{Astronomy and Astrophysics Division, Physical Research Laboratory,\\
Navrangpura, Ahmedabad - 380 009, India}
\email{asingal@prl.res.in} 
\begin{abstract}
In the currently popular orientation-based unified scheme, a radio galaxy appears as a quasar 
when its principal radio-axis 
happens to be oriented within a certain cone opening angle around the observer's line of sight. 
Due to geometrical projection, the observed sizes of quasars 
should therefore appear smaller than those of radio galaxies.  
We show that this simple, unambiguous prediction of the unified scheme is not borne 
out by the actually observed angular sizes of radio galaxies and quasars. 
Except in the original 3CR sample, based on which the unified scheme was proposed, 
in other much larger samples no statistically significant difference is apparent in 
the size distributions of radio galaxies and quasars. The 
population of low-excitation radio galaxies with apparently no hidden quasars inside,
which might explain the observed excess number of radio galaxies at low redshifts, cannot 
still account for the absence of any foreshortening of the sizes of quasars at large redshifts. 
On the other hand from infrared and X-ray studies there is evidence of hidden quasar within a 
dusty torus in many RGs, at $z>0.5$. It seems difficult how to reconcile this with the absence of foreshortening 
of quasar sizes at even these redshifts, and perhaps one has to allow that the major radio axis may not have 
anything to do with the optical axis of the torus. Otherwise to resolve the dichotomy of radio galaxies and quasars, 
a scheme quite different from the present might be required.
\end{abstract}
\section{Introduction}
The observed numbers and radio sizes of quasars both 
appear to be about a factor of two smaller than those of radio galaxies (RGs), in the radio strong 
3CR complete sample (Laing et al. 1983), in the redshift range $0.5<z<1$ (Barthel 1989).  
It was suggested that both RGs and quasars belong to the same
parent population of radio sources, and that a source appears as
a quasar only when its principal radio-axis happens to be oriented within a certain cone 
opening angle ($\xi_{c}$) around the observer's line of sight (Barthel 1989).
In this model, the nuclear continuum and broad-line optical emission region 
is surrounded by an optically-thick torus and $\xi_{c}$ is the half cone-opening  
angle of the torus, similar to as proposed in the case of Seyfert galaxies (Antonucci \& Miller 1985). 
In the case of RGs the observer's line of sight is supposed to be 
passing through the obscured region which hides the bright optical nucleus and the broad-line region. 
Accordingly, RGs and quasars are considered to be intrinsically indistinguishable 
and all differences in their observed radio properties  are 
attributed to their supposedly different orientations with respect
to the observer's line of sight; in particular, the observed smaller value
of radio sizes of quasars in the 3CR sample was attributed to their larger
geometric projection effects because of the shallower inclinations
of their radio axes with respect to the observer's line of sight.

This has come to be known as orientation-based unified scheme (OUS) and has gained increasing
popularity (Antonucci 1993; Antonucci 2012; Urry \& Padovani 1995; Kembhavi \& Narlikar 1999) 
both because of its simplicity and the promise it holds to bring two apparently 
quite distinct class of objects, viz. quasars and RGs, under one roof. 
According to this scheme, the expected ratios of the observed numbers as well as of 
sizes of quasars and RGs in a low-frequency radio-complete sample
are determined purely by the value of $\xi_{\rm c}$.
It is widely believed that, in samples picked at metre wavelengths, the
observed number as well as sizes of quasars are typically about half as large as those
of RGs. This notion has resulted purely from the data in a limited redshift range 
($0.5<z<1$) of the 3CR sample that yielded the `canonical' value 
of $\xi_{\rm c} \sim 45^{\circ}$. Later  Singal (1993a) pointed out 
that the data in other redshift bins from the rest of the 3CR sample do not seem to 
fit into this simple scenario. Suggestions were then put forward 
(Gopal-Krishna et al. 1996) that by making allowance for a temporal evolution 
of sources in both size and luminosity, one could mitigate the above discrepancy. 
Alternatively it has been suggested that 
this excess may be due to a population of low-excitation radio galaxies (LERGs),
which might make  a significant contribution to the number of FR~II-type radio galaxies 
at low redshifts (see e.g. Hine \& Longair 1979).
Laing et al. (1994) have pointed out that these optically dull LERGs are unlikely to appear
as quasars when seen end-on and that these should be excluded from the sample while testing the unified 
scheme models. From Infra-red observaions also there is evidence of a population of powerful radio galaxies, concentrated at 
low redshifts, which lack the hidden quasar (Antonucci 2012; Ogle et al. 2006; Leipski et al. 2010). 
Using both X-ray and Mid-IR data, Hardcastle et al. (2009) showed conclusively that almost all objects classed as LERGs
in optical spectroscopic studies lack a radiatively efficient active nucleus.   
On the other hand strong evidence against OUS comes also from the observed opposite behaviour of the 
luminosity--size correlations among RGs and quasars as well as from the vast difference in their cosmological 
size evolutions (Singal 1988, 1993b, 1996a).

A comparison of the angular size of RGs and quasars is a very robust test, as in samples
selected at metre wavelengths, emission only from the steep spectrum extended parts of the source is observed,
with flat-spectrum core-emission, if any, highly suppressed and the relativistic beaming effects 
playing almost no part. Both quasars and RGs are picked by the strength of their extended emission, 
not affected by any orientation effects. 
As the parent sample for RGs and quasars is supposed to be the same, 
there will be no relative selection effects based on redshift or luminosity, with 
their observed size ratios affected only by the geometrical projection. 
Further, it is not necessary to convert their angular sizes into linear sizes (using a 
particular cosmological model) for the comparison of their sizes to test OUS as the observed angular size ratios will 
truly reflect their (projected) linear size ratios since the redshift distribution is supposed to be the same for 
RGs and quasars in OUS models. If we think that the redshift distribution might be different for RGs 
and quasars, then we are already doubting the veracity of the unified scheme. 
\section{The Source Sample}

For our investigations we have chosen an essentially complete MRC sample (Kapahi et al. 1998a,b), 
which is about a factor $\sim 5$ deeper than the 3CR sample and has the required radio and optical information. 
It comprises a total of 550 sources, with 105 of them being quasars, six BL Lac objects,   
and the remainder RGs. Optical identifications for the latter are complete up to a red magnitude of $\sim 24$ 
or a {\em K} magnitude of  $\sim 19$. Spectroscopic redshift data are available for 60 percent of the galaxies, 
the remainder are mostly faint galaxies (McCarthy et al. 1996) expected to be at high redshifts $z \stackrel{>}{_{\sim}}1$.
Optical spectroscopic data for quasars with full observational details are given in Baker et al. (1999),  
with  tabulations of redshifts, continuum, and emission-line data for each source. 
The optical identifications are missing only for a very small percent of sources (Kapahi et al. 1998a), 
which should not be too detrimental for our investigations here. 

As only the powerful RGs are supposed to partake in unification with quasars, 
we have confined ourselves to only the strong, FR~II-type sources (Fanaroff \& Riley 1974) 
with $P_{408} \ge 5 \times 10^{25}$ W Hz$^{-1}$ 
(for a Hubble constant $H_{0}=71\,$km~s$^{-1}$\,Mpc$^{-1}$, the matter energy density $\Omega_m=0.27$, and 
the vacuum energy (dark energy!) density $\Omega_{\Lambda}=0.73$; Spergel et al. 2003);  
the quasars in any case (all but one) fall above this luminosity limit. 
This limit corresponds to the FR I/II luminosity break  
$P_{178} = 2 \times 10^{25}$ W Hz$^{-1}$ sr$^{-1}$ (for $H_{0}=50\,$km~s$^{-1}$\,Mpc$^{-1}$) 
of Fanaroff \& Riley (1974). It may be prudent to exclude 
flat-spectrum sources entirely, since these mostly are core-dominant cases where
the relativistic beaming might introduce serious selection effects. 
Among the quasars there are 16 sources with spectral index $\alpha \le 0.5$ (with $S\propto \nu^{-\alpha}$), while 
among the RGs there are only 7 such cases; we have excluded all these flat-spectrum cases. 
Also there is a large fraction of Compact Steep Spectrum Sources (CSSS; linear size $< 25$ kpc) in the MRC sample, 
comprising about 20 percent of the whole sample, which seem to be a different class  
than the FRII class of sources whose unification is sought in OUS (Kapahi et al. 1995), and as such 
these should be excluded for testing OUS. For the unknown-$z$ galaxies we have taken $<3$ arcsec as the CSSS 
criteria as at $z \sim 1$, where most of these faint galaxies are likely to be, for our adopted cosmological parameters, 
1 arcsec translates to about 8 kpc. 
 
It has been pointed out (Kapahi et al. 1995) that radio sizes of MRC quasar appear considerably larger than 
those of 3CR quasars. Such can of course be taken only as an indicator of some anomaly, 
as the flux-density range and the volumes sampled in the two samples (MRC and 3CR) are very different. 
For a meaningful comparison of the relative sizes of RGs and quasars, in order to test OUS, 
one must draw both kind of sources (RGs and quasars) from a common parent sample, so that no effects enter 
due to different flux-density and/or the space-volume sampled as any redshift or luminosity dependence of the sizes could 
otherwise bias the conclusions. Earlier it could not be done because data on the MRC sample of galaxies were then not 
complete (Kapahi et al. 1995), but these data having since become available, we could now attempt it here. 

There are a total of 494 sources in our sample listed in Table 2, which is organized in the
following manner: 
(1) Source name from MRC.
(2) Flux-density $S_{408}$ at 408 MHz.
(3) Spectral index $\alpha$ ($S\propto \nu^{-\alpha}$).
(4) Nature of optical object; G: galaxy; Q: quasar; U: unidentified.
(5) Redshift $z$, whenever a measured value is available.
(6) Largest angular size $\theta$ (in arcsec). 
(7) Linear size $l$ in kilo-pc.
(8) Luminosity $P_{408}$ in W/Hz.
Among these 494 sources, there are 379 RGs, 87 quasars, and 28 remain unidentified.
The linear size is calculated from the observed angular size 
$\theta$ as $l=\theta {\cal D}/(1+z)$ and the luminosity is calculated from $P_{408}=4 \pi S_{408}{\cal D}^2(1+z)^{1+\alpha}$, 
where ${\cal D}$ is the comoving cosmological distance calculated from the cosmological redshift $z$ of the source. 
In general it is not possible to express ${\cal D}$ in terms of 
$z$ in a close-form analytical expression and one may have to evaluate it 
numerically. For example, in the  flat universe models   
($\Omega_m+\Omega_\Lambda=1, \Omega_\Lambda \neq 0$), ${\cal D}$ is given by (see e.g., Weinberg 2008),
\begin{eqnarray}
{\cal D}=\frac{c}{H_0}\int^{1+z}_{1}\frac{{\rm d}z}{\left(\Omega_\Lambda+\Omega_m z^3\right)^{1/2}}\;.
\end{eqnarray}
For a given $\Omega_\Lambda$, ${\cal D}$ can be evaluated from Eq. (1) by a numerical integration. 
\begin{figure}[ht]
\scalebox{0.4}{\includegraphics{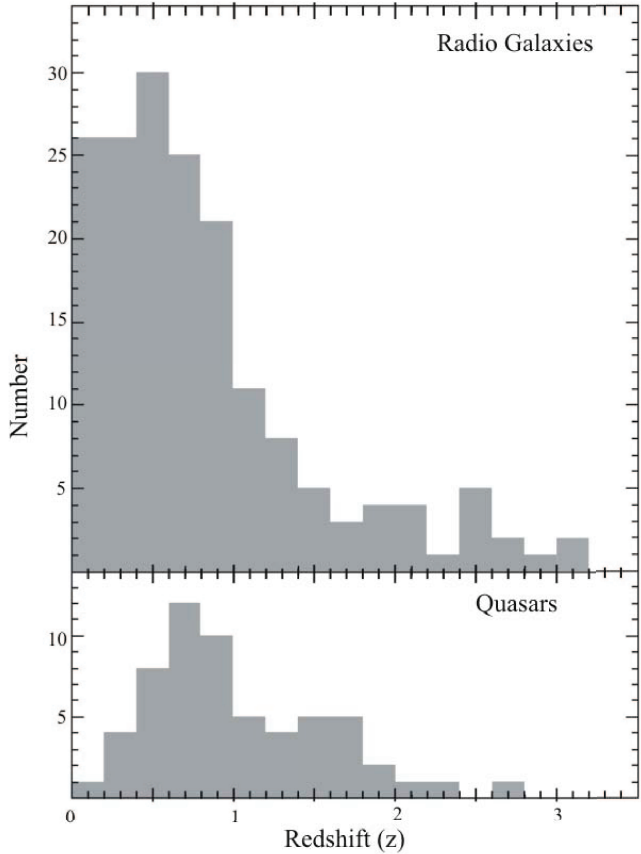}}
\caption{Histograms of the redshift distributions of RGs and quasars for the MRC sample.}
\end{figure}
\begin{figure}[ht]
\scalebox{0.4}{\includegraphics{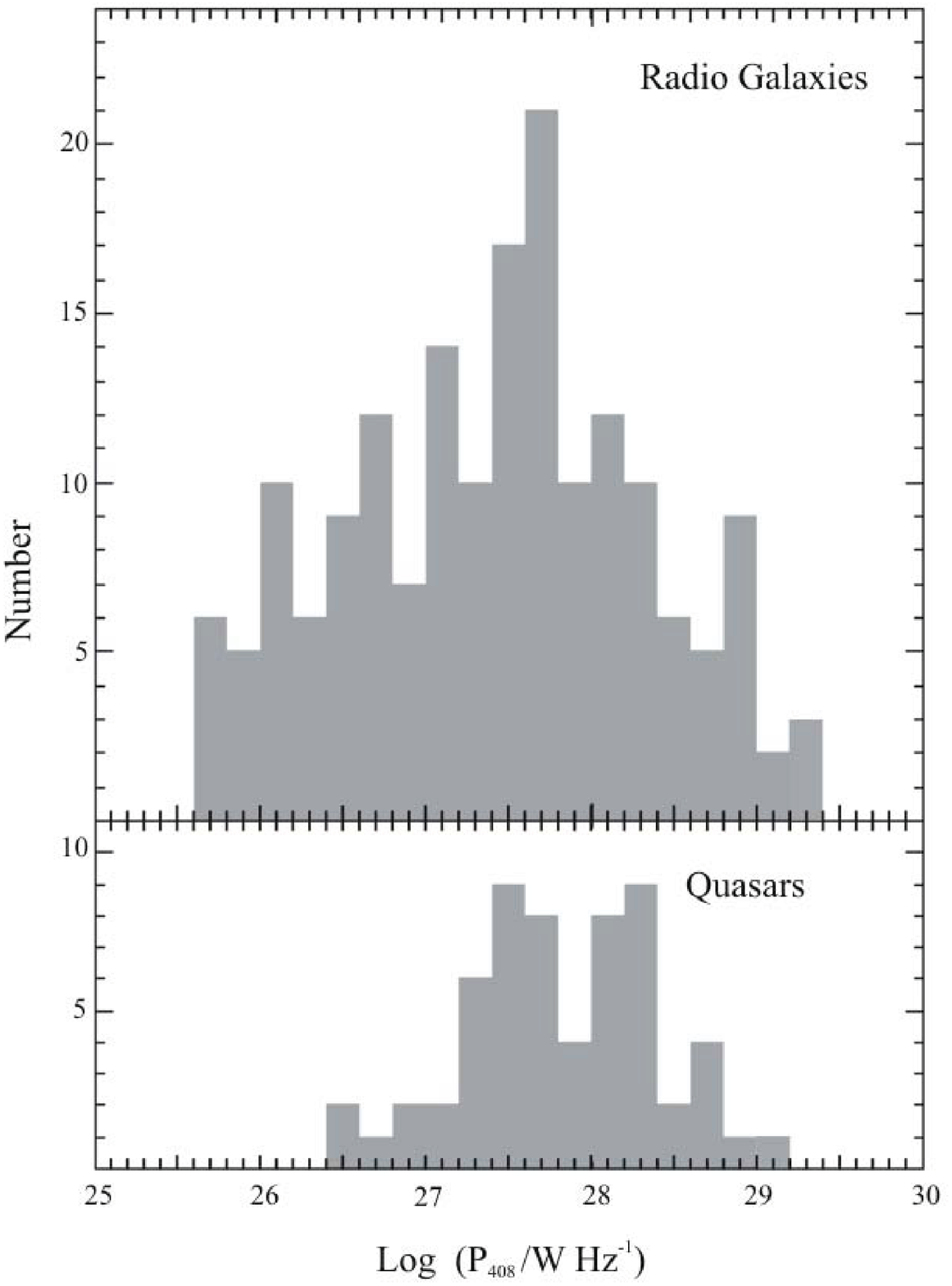}}
\caption{Histograms of the luminosity distributions of  RGs and quasars for the MRC sample.}
\end{figure}
\section{Results and Discussion}
Figure 1 shows the redshift distribution of RGs and Quasars in our sample. We notice a large excess of 
RGs at low redshifts ($z\stackrel{<}{_{\sim}}0.5$), similar to the excess seen in the 3CR sample, as   
pointed out by Singal (1993a). Figure 2 shows the luminosity distribution of RGs and Quasars. 
Again we notice an excess of RGs at lower end of the FRII luminosities 
($P\stackrel{<}{_{\sim}} 10^{27}$ W Hz$^{-1}$). Of course we expect the luminosity distribution to largely mimic the redshift 
distribution because of the Malmquist bias in a flux-limited sample like the MRC. 
It also needs to be noted that the unknown-$z$ RGs in the MRC sample will mostly fall at high ends of redshift and luminosity;  
in any case we are already finding a rather surplus number of RGs at low redshifts and luminosities than that expected from OUS.
\begin{figure}[ht]
\scalebox{0.55}{\includegraphics{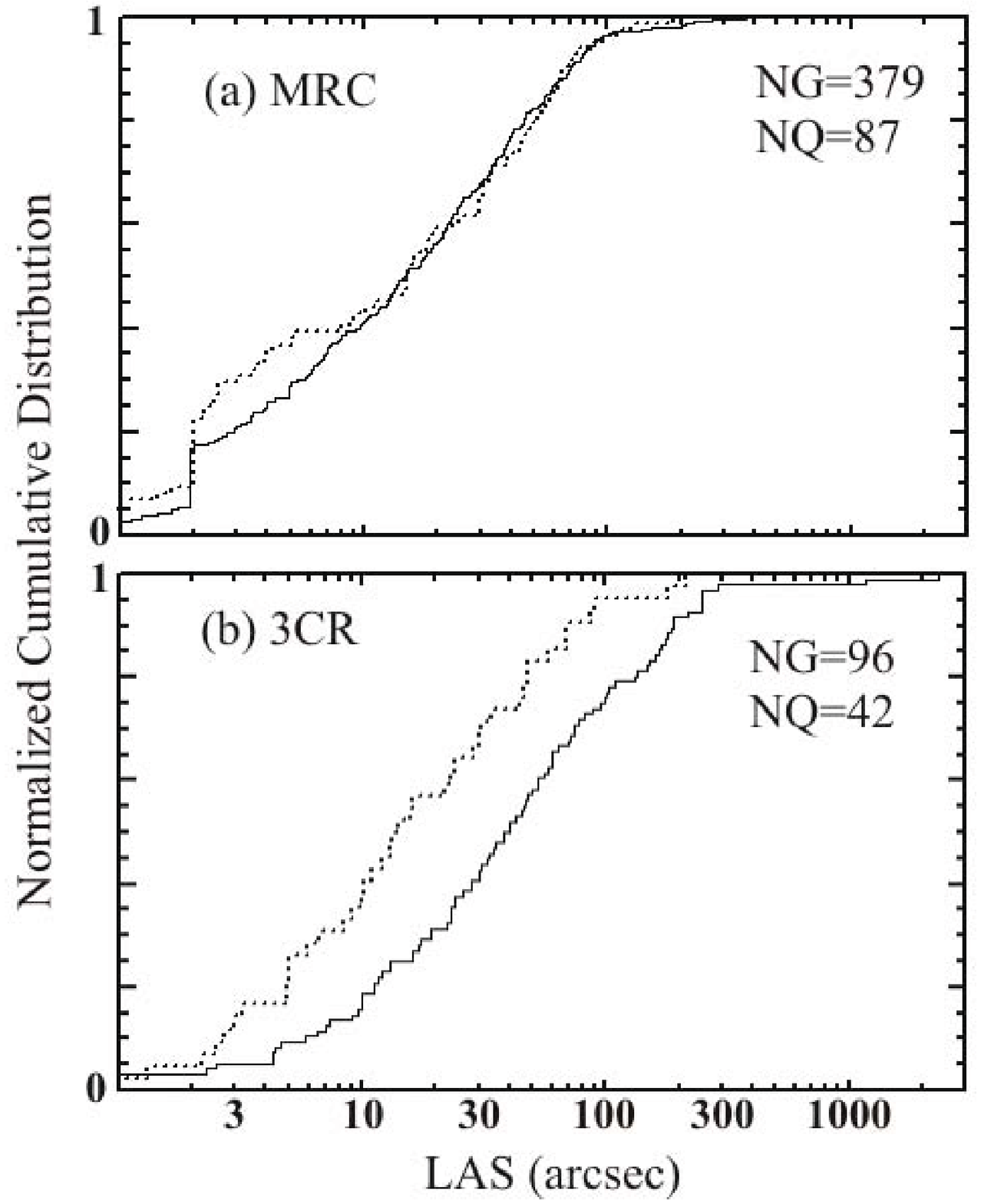}}
\caption{Normalized cumulative distributions of largest angular size (LAS) of RGs (continuous curves) and 
quasars (broken curves) (a) for the MRC sample (b) for the 3CR sample. NG and NQ give the number of RGs 
and quasars respectively, in each plot.}
\end{figure}

As mentioned earlier LERGs, a population of FR II RGs with no hidden quasars, concentrated only at low redshifts (say, $z<0.5$)  
could make the apparently anomalous number and size distribution of RGs and quasars at low redshifts in the 3CR sample 
(Singal 1993a) somewhat consistent with OUS, and perhaps it might also hold true for the MRC sample. 
For that more than half of the total source population at these redshifts will have to be LERGs. Another implication will
be that LERGs, if they do indeed form an isotropic sample (as suggested by Laing et al. 1994), should show smaller projected
sizes as compared with those of high-excitation radio galaxies, which
supposedly lie preferentially in the sky plane. There is some evidence for that (Hardcastle et al. 1998).
But could such a population of LERGs be also present in the low freqency samples at higher redshifts? 
For one thing such a scenario would imply that a large majority of FR~II radio galaxies ($\sim 50-60 \%$) remain a 
separate class (with intrinsically  different properties from quasars/BLRGs) and not fall within the scope of OUS.
At the same time it is also clear that if such a high percentage of LERGs is 
making up the low freqency samples like the 3CR at higher redshifts $z\stackrel{>}{_{\sim}} 0.5$ as well, then 
the number and size ratios, used by Barthel in his original 3CR sample in the redshift range $0.5<z<1$ to propose OUS, 
will totally go haywire and the proposition of OUS will have to be abondoned in the first place there itself.
\begin{figure}[ht]
\scalebox{0.7}{\includegraphics{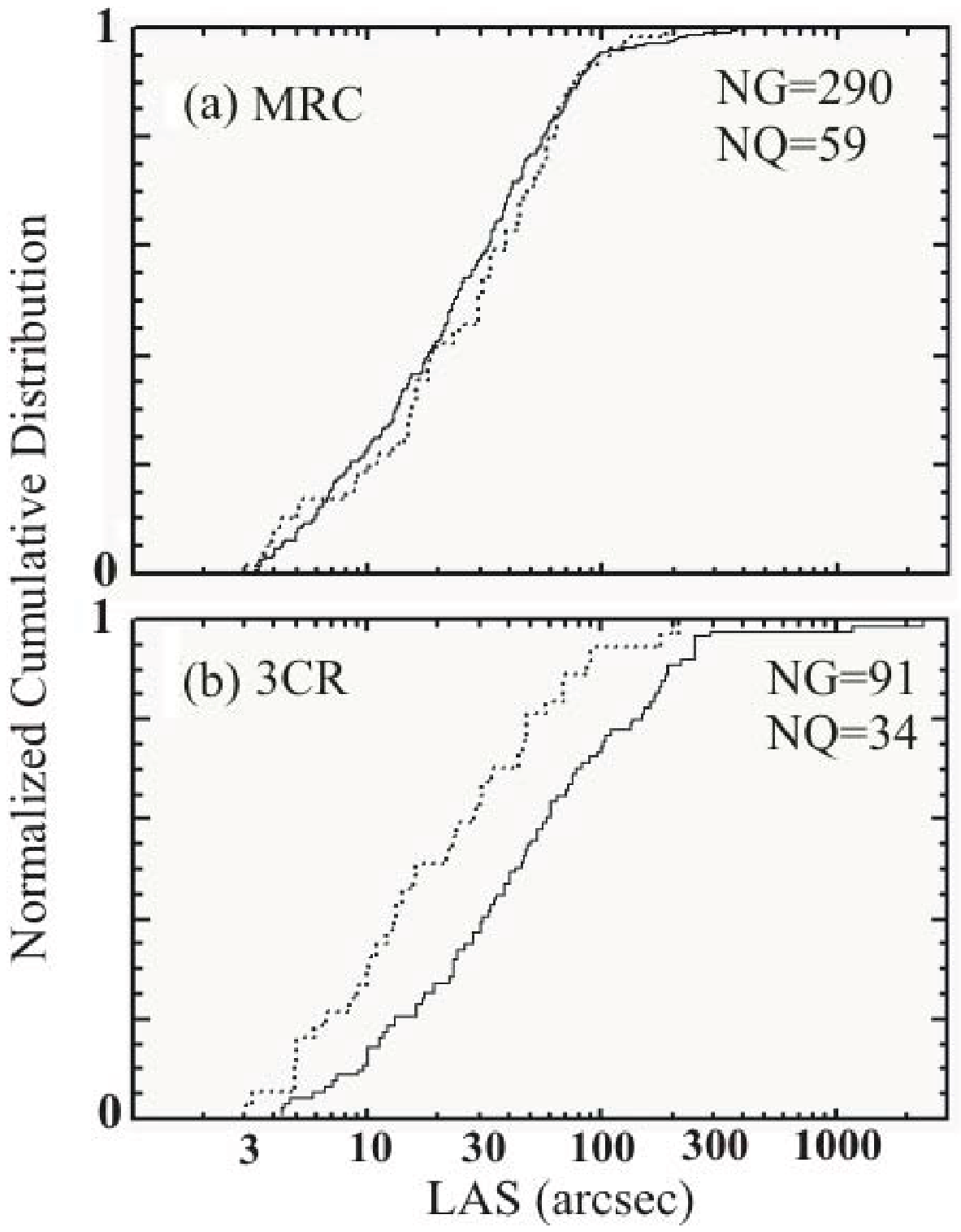}}
\caption{Same as Figure 3, but with CSSS excluded.}
\end{figure}

Due to geometrical projection, the observed sizes of quasars should appear smaller than those of radio galaxies.  
This is a simple, unambiguous prediction of the unified scheme which is thus falsifiable from a comparison of the 
observed angular sizes of radio galaxies and quasars. 
We therefore examine OUS by this robust test of the relative size distributions of RGs and quasars. 
Figure 3a shows normalized cumulative plots of angular size distribution of RGs and quasars 
for our chosen MRC sample, while for a comparison Figure 3b show the same for the 3CR sources.
In the 3CR sample, as expected, the quasar sizes seem smaller than those of the RGs which of course was the prime basis for  
the OUS hypothesis. However in a much larger independent MRC sample, there seems no evidence that  
the quasar sizes are in any way smaller than those of RGs. 
Here we had retained the CSS sources and we notice a bump (discontinuity!) at or around 2 arcsec in the MRC size 
distributions both for RGs and quasars, due to these CSS Sources. 
But in figure 4, we have excluded all the CSSS cases and we see that the inclusion or exclusion of CSSS in either 
case does not alter any of our conclusions. 

Since the two samples (MRC and 3CR) have different flux-density limits, it may be interesting to see if the 
difference between the two samples in the size distributions is in any way related to the flux-density level of the samples. 
Figure 5 shows the normalized cumulative distribution of radio sizes of RGs and quasars  
for the MRC sample in three different flux-density bins. To avoid any selection bias, we have chosen the 
flux-density bins such that there are about equal 
number of sources in each bin. We see no change in the earlier picture, the quasar and RGs do not   
show any systematic difference in their size distributions. 
With its basic tenet, that the observed quasar sizes should be smaller than of RGs,   
having been precluded, OUS is thus almost ruled out. 
\begin{figure}[ht]
\scalebox{0.4}{\includegraphics{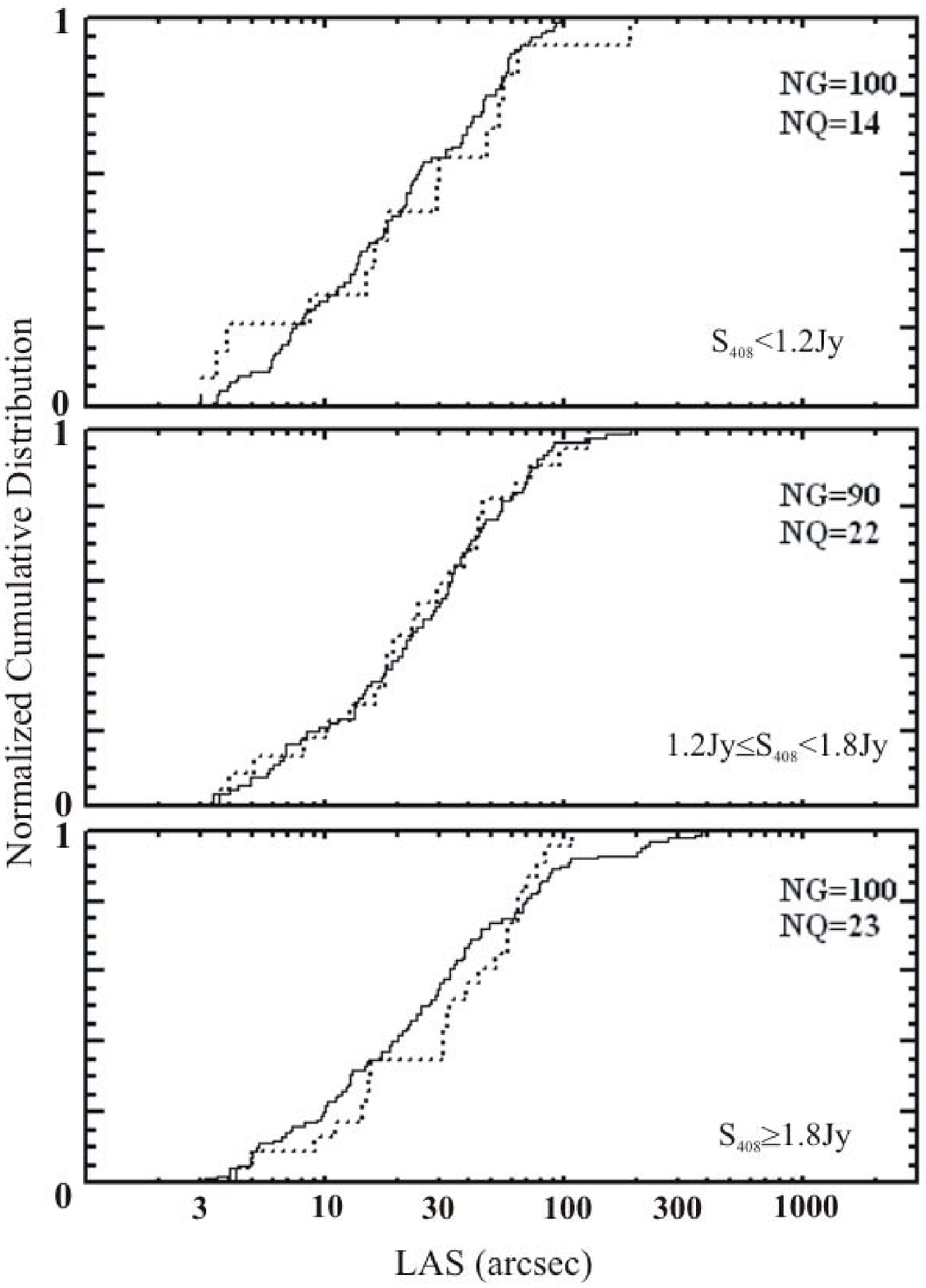}}
\caption{Normalized cumulative distribution of radio sizes of RGs (continuous curves) and quasars (broken curves)
for the MRC sample in different flux-density bins.}
\end{figure}

We should clarify that the evidence against the unification of extended RGs and quasars here does not necessarily 
invalidate the relativistic beaming models (Orr \& Browne 1982) of the unification of core-dominated and 
lobe-dominated quasars. In the same way, any evidence seen 
in favour of the relativistic beaming models cannot be cited in favour of unification of extended RGs and quasars. 
The two unifications are independent even if these have been combined in the so-called grand unification  
scheme models of the active galactic nuclei (Antonucci 1993; Antonucci 2012; Urry \& Padovani 1995; 
Kembhavi \& Narlikar 1999). It  is quite likely that radio-loud quasars do not make a randomly oriented population;  
the question here is that do RGs and quasars fit together, as proposed by Barthel (1989), 
into one unified scheme model like OUS?
\begin{figure}[ht]
\scalebox{0.4}{\includegraphics{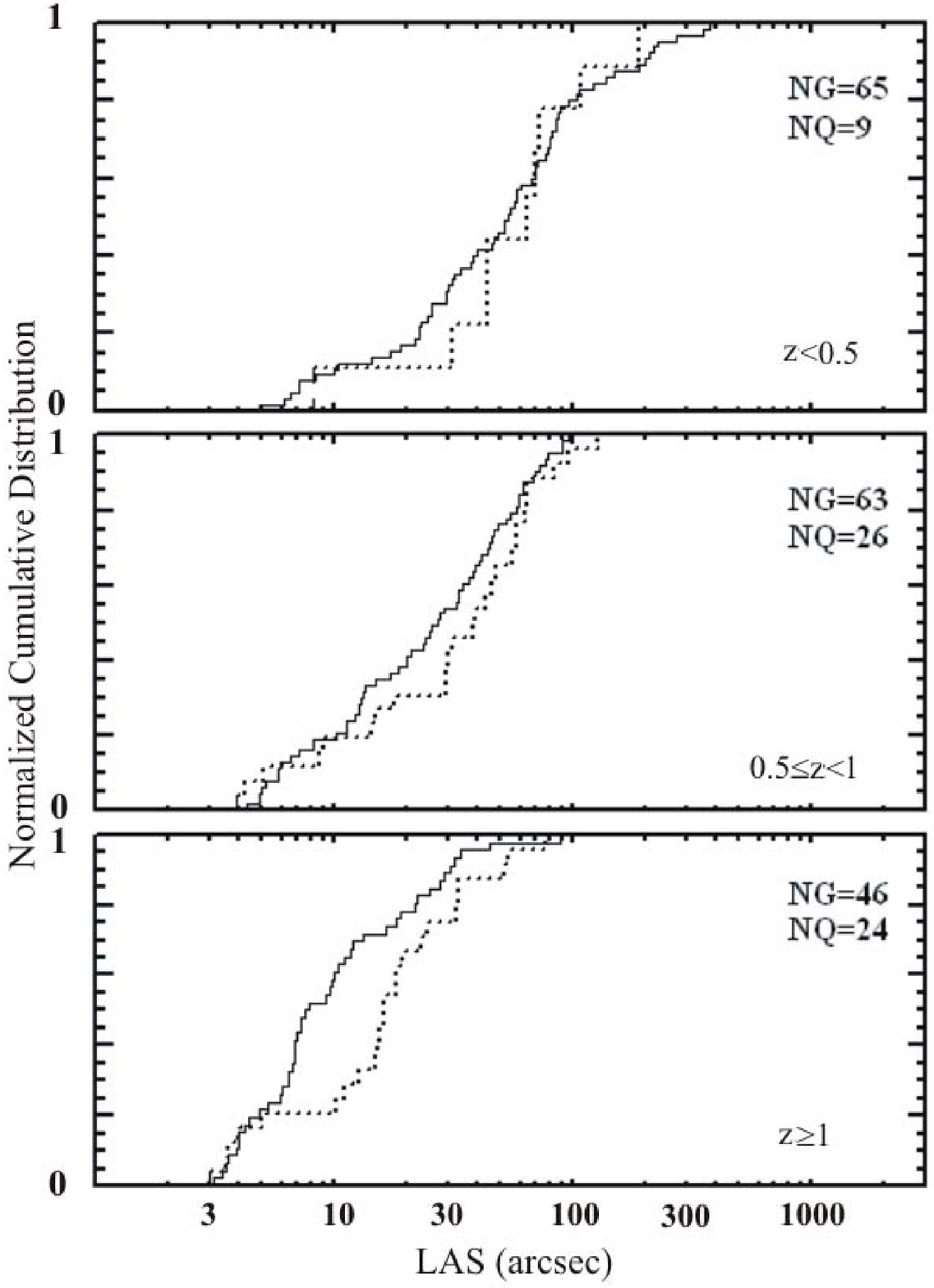}}
\caption{Normalized cumulative distribution of radio sizes of RGs (continuous curves) and quasars (broken curves) 
for the MRC sample in three different redshift ranges.}
\end{figure}

We compare in Figure 6 the size distributions of MRC sources in three different redshift bins. 
Again we find no evidence for quasars being smaller in size in any of the three redshift bins. 
As we have 116 RGs withour redshift determination, and most of these will be at $z\stackrel{>}{_{\sim}} 1$, 
in Figure 7a we include these along with 46 RGs with $z \geq 1$ to compare the angular sizes of all 
these high redshift RGs with those of quasars at $z \geq 1$. We find that the size distributions are strikingly similar. 
There are also 28 unidentified cases, which again most likely will be RGs at $z\stackrel{>}{_{\sim}} 1$. 
After dropping 9 CSSS cases among them, we compare in Figure 7b the size distribution of 19 unidentified cases 
with that of quasars with $z \geq 1$. The two size distributions statistically look almost indistinguishable. 
\begin{figure}[ht]
\scalebox{0.55}{\includegraphics{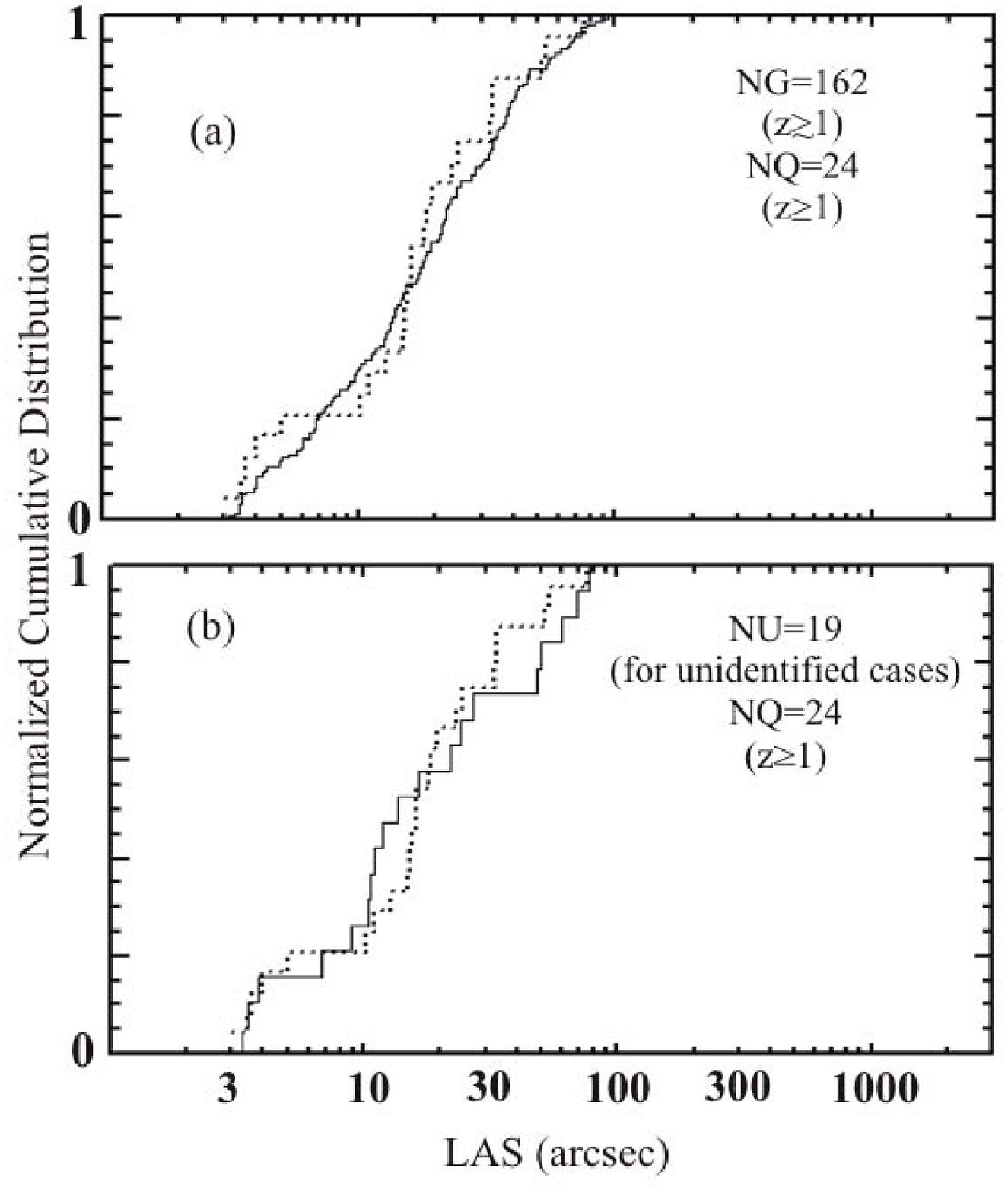}}
\caption{Normalized cumulative distribution of radio sizes of (a) RGs (continuous curves) comprising 46 sources with $z\geq 1$  
and 116 unknwon--redshift galaxies (a total of 162 RGs with expected $z\stackrel{>}{_{\sim}} 1$), 
and quasars (broken curves) for $z\geq 1$ and (b) unidentified sources (continuous curves) and quasars (broken curves) 
for $z\geq 1$.}
\end{figure}

We have determined median value $\theta_{med}$ (in arcsec)
of the cumulative size--distribution of the source in the various sub-samples (Table 1). 
To get an idea of the spread around the median values we have also listed in Table 1 the lower quartiles $\theta_{\rm lq}$ 
and upper quartiles $\theta_{\rm uq}$ in all cases.
We find that while in the 3CR sample quasar sizes may be half those of RGs, in the much larger MRC sample the 
roles seems to have been reversed (cf. Table 1) with quasars appearing to be in fact somewhat larger in size than RGs 
(in the whole MRC sample by a factor of $\sim 1.3$ though in a sub-sample like $z\geq 1$ by as much as a factor of two); 
in any case nowhere is there an evidence of the quasar sizes being smaller than of RGs. This is apparent not only 
from median values of size but also from the lower and upper quartile values in their cumulative plots. 
\begin{table*}[ht]
\caption{Median and quartile values of size distributions for RGs and quasars.}
\begin{tabular}{@{\hspace{0mm}}c@{\hspace{3mm}}c@{\hspace{3mm}}c@{\hspace{3mm}}c@{\hspace{3mm}}
c@{\hspace{10mm}}c@{\hspace{3mm}}c@{\hspace{3mm}}c@{\hspace{3mm}}c}
 & & & & & & & & \\
Sub-sample & NG & $\theta_{\rm lq}$ & $\theta_{\rm med}$ & $\theta_{\rm uq}$ & 
NQ & $\theta_{\rm lq}$ & $\theta_{\rm med}$  & $\theta_{\rm uq}$  \\
 & & & & & & & & \\
All 3CR & 91 & $17$  & 40& 98 & 34 &  $10$  & 18& 45    \\\\
All MRC & 290 & $11$  & 22& 46 & 59 &  $14$  & 30& 54    \\
 & & & & & & & & \\
$S_{408} < 1.2$ & 100 & $9$  & 21& 42  & 14 & 8  & 24 &  50 \\
$1.2\leq S_{408}<1.8$ & 90 & $14$ & 26 &   45 & 22 & $12$ & 24 &   44 \\
$S_{408}\geq 1.8$ &  100 & $12$ & 27& 55 & 23 & $15$ & 33 &  60\\
& & & & & & & & \\
$z < 0.5$ & 65 & $25$  & 54& 86  & 9 & $38$  & 64 &  91 \\
$0.5\leq z<1 $ & 63 & $11$ & 27 &   38 & 26 & $15$ & 39 &   58 \\
$z\geq 1$ &  46 & $6$ & 8& 17 & 24 & $11$ & 16& 24\\
\end{tabular}
\end{table*}

The cone opening angle ($\xi_{c}$) consistent with the smaller quasar fraction seen in MRC data (Figures 4, 5 and 6; 
Table 1) will mean a value narrower than $\sim 45^{\circ}$ derived from the 3CR data, implying expected size ratio to 
be even more pronounced than that in the 3CR case. 
After dropping CSS sources, in the MRC sample the fraction of quasars is $59/(290+59) \sim 0.17$, implying only about 
one sixth of the sources 
are quasars (as compared to one third in 3CR sample of Barthel (1989), based on which OUS, with a cone opening angle 
$\xi_{c}\sim 45^{\circ}$, was proposed). That such a low fraction of quasars is in itself an evidence against the popular OUS scheme was 
already pointed out by Singal (1996b). In a picture consistent with OUS, MRC quasar sizes should be statistically smaller 
than of RGs by more than a factor of two, but we on the other hand find quasars to be rather somewhat bigger in size 
(by a factor of $\sim 1.3$) than even of RGs, which could never happen in an OUS type of scheme. 
Thus there is no consistency at all in the number count ratios and the size ratios and no cone opening angle ($\xi_{c}$) can be 
found within OUS that would satisfy both relative number counts and relative size distributions observed for both quasars and RGs. 
Thus the predictions of OUS are not corroborated by the data in a sample other that the 3CR at even high redshift bins 
$z>0.5$ (and at high luminosities), where LERGs may not play an important part, and OUS is clearly ousted in that sense.

It is clear that at a few Jy or weaker levels, 
OUS does not hold good. If we still want to hold on to OUS, in the belief  
that it might be valid at or above only the higher flux levels of 3CR radio sources ($S_{408}>4$ Jy), then since the 
integrated source counts fall rapidly with flux ($N(>S) \propto S^{-1}$), it is only a tiny 
fraction of the RGs that would be taking part in the currently popular OUS. 
In fact one would then be proposing a division of RGs into further sub-classes (beyond LERGs seen only at 
low redshifts and low luminosities within FRII types RGs), out of which 
perhaps only one minor sub-class of RGs will be partaking in OUS, thus sounding more like a further 
{\em disunification scheme} for RGs. And even there it will have to be one of those ``cosmic conspiracies'' 
where two or more sub-classes of RGs of apparently very different size distributions manage to get their 
combined size-distribution very similar 
to that of quasars in all flux-density and redshift bins.

By still adhering to the belief that may be a restricted class of RGs partakes in the 
OUS model, it seems that much effort is being put on giving perhaps rather undue weight to a small select 
sub-sample ($0.5<z<1$ redshift bin of 3CR) that happened to be the first one to get examined in this regard. 
Except for that particular bin of the 3CR sample, which incidentally was instrumental in the proposition of the unified scheme 
with the ``canonical'' value $\xi_{\rm c}\sim 45^{\circ}$, other samples do not seem to yield the expected size ratios, 
in fact as we saw there seems to be no statistically significant difference in the size distribution of quasars and RGs.

At the same time infrared and X-ray studies do find many cases of obscured hidden quasar in powerful radio galaxies, 
implying that a unified scheme must be true to some extent.
Most 3CR FRII radio galaxies with $0.5\stackrel{<}{_{\sim}} z<1$ do show strong, quasar-like mid-IR emission (Antonucci 2012), 
while  at $z\stackrel{<}{_{\sim}} 0.5$ there is suggestion of a dearth of hidden AGN in FRII RGs. For example, 
Meisenheimer et al (2001) observed at infrared wavelengths 10 quasars and 10 RGs, selected 
with matched luminosity and the redshift distributions from the 3CR catalogue, and found the results compatible with 
hidden Quasars, except possibly for some at the low luminosity/redshift end.
Shi et al (2005) used the Spitzer photometry to study a sample of 3CR radio galaxies and 
the behavior of the sources is consistent with the presence of an obscuring circumnuclear torus.
The X-ray spectroscopic survey of 38 high-z 3CR objects of Wilkes et al (2009) 
is extremely supportive of complete unification of 3CR radio galaxies and Quasars at $z\stackrel{>}{_{\sim}} 1$. 
It seems rather quizzical how to reconcile such overwhelming evidence for torus with the equally strong evidence for 
the lack of foreshortening of radio sizes of quasar at all redshifts. 

The predictions of the unified scheme models 
are not corroborated by the radio observations and that seems to refute the 
presently popular OUS. At the same time it does not seem likely that, except perhaps in some very contrived scenario,
any modification of OUS, such as an evolutionary model of $\xi_{\rm c}$ with redshift or luminosity, could yield quasar 
sizes comparable to those of galaxies in all bins, since in OUS those will be expected to be smaller due to geometrical 
projection everywhere. On the other hand quasar sizes are not found to be smaller than those of galaxies  
for any of the bins, whether in flux-density or in redshift, in the MRC sample. Perhaps we may have to allow that the orientation 
of the extended radio structure does not relate to the axis of the torus, which amounts to abandoning a basic tenet of OUS, or
we may require some very different unification scheme than the currently popular orientation-based unified scheme model. 
\section{Conclusion}
We showed that contrary to the expectations in OUS models, observed quasar sizes are not in any way systematically smaller than 
those of galaxies. The absence of this foreshortening of the sizes of quasars as compared to those of RGs of similar flux densities 
or at similar redshifts, provides irrefutable evidence against the unified scheme models. 
To still uphold OUS, one would need to propose FRII type RGs with no hidden quasars, and/or of small intrinsic radio sizes 
at {\em all} redshifts and luminosities, i.e., across the entire gamut of population of strong radio galaxies, 
with a large majority of RGs opting out of OUS. It means first one would rather require a robust 
{\em disunification scheme} of the radio galaxies themselves before an attempt could be made 
to unify a rather small number of RGs with quasars in a scheme like OUS. Or perhaps one has to allow that the major 
radio axis does not coincide with the axis of the torus, which is a basic tenet of the present unification scheme.
In any case it appears that the dichotomy of RGs and quasars among 
extragalactic radio sources cannot be resolved within the present frame-work of OUS, which 
therefore seems falsified. 

\clearpage
\begin{longtable}{cllclrrc}
\caption{Radio and optical data for our sample.}\\
Source & $S_{408}$ & $\,\, \alpha$ & Opt & $\,\,z$ & $\theta\,\, $ &  $\l\,\,\, $ & $P_{408}$ \\
Name & \,\,Jy & & Obj & & \arcsec\,\,  & kpc & W Hz$^{-1}$\\
(1)&\,(2)&(3)&(4)&(5)&(6)&(7)&(8)\\
\\
\endfirsthead
\tablename ~2\ -- \textit{continued} \\
Source & $S_{408}$ & $\,\, \alpha$ & Opt & $\,\,z$ & $\theta\,\, $ &  $\l\,\,\, $ & $P_{408}$ \\
Name & \,\,Jy & & Obj & & \arcsec\,\,  & kpc & W Hz$^{-1}$\\
(1)&\,(2)&(3)&(4)&(5)&(6)&(7)&(8)\\
\endhead
B0001-237 &1.77&0.83&G&0.315&33.8&155&5E26\\
  B0006-212 &1.18&0.73&G&0.91&0.9&7&4E27\\
  B0007-287 &1.13&0.87&U& &68.3& & \\
  B0015-229 &1.11&1.16&G&2.01&10.5&89&4E28\\
  B0017-205 &1.96&0.78&G&0.197&372&1202&2E26\\
\\
   B0017-207 &1.25&0.95&Q&0.545&96&612&1E27\\
  B0020-253 &5.36&0.78&G&0.35&79&388&2E27\\
   B0022-297 &7.83&0.87&Q&0.406&44&238&4E27\\
  B0023-203 &3.43&0.95&G&0.845&8.1&62&1E28\\
  B0023-263 &17&0.64&G&0.322&1.9&9&5E27\\
\\
  B0025-204 &1.03&0.77&G& &11.1& & \\
  B0025-277 &1.29&0.92&G& &13.7& & \\
  B0028-223 &1.18&0.79&G&0.205&23.2&77&1E26\\
  B0029-232 &1.09&1.07&G& &21.9& & \\
  B0029-243 &1.96&1.15&G&1.29&4&34&2E28\\
\\
   B0029-271 &1.21&0.96&Q&0.333&1&5&4E26\\
  B0030-219 &1.08&1.06&G&2.168&0.9&8&4E28\\
   B0030-220 &1.01&0.93&Q&0.806&3.9&29&3E27\\
  B0030-297 &1.63&1.02&G& &1.9& & \\
  B0032-203 &6.87&1.08&G&0.516&1.5&9&7E27\\
\\
  B0034-234 &1.69&1.11&G& &15.6& & \\
  B0035-231 &1.21&0.99&G&0.685&2.3&16&2E27\\
  B0037-258 &1.21&0.96&G&1.1&27.6&227&8E27\\
  B0038-294 &1.01&0.87&G& &17.5& & \\
   B0040-208 &1.12&0.9&Q&0.657&2&14&2E27\\
\\
  B0041-224 &1.85&1.23&G& &45& & \\
  B0042-248 &1.45&0.87&G& &73& & \\
  B0050-222 &1.41&0.98&G&0.654&1.9&13&3E27\\
  B0052-241 &1.18&1.14&G&2.86&2.5&20&1E29\\
  B0055-256 &1.18&0.97&G&0.199&22.8&74&1E26\\
\\
  B0055-258 &1.02&1.13&G& &58& & \\
  B0056-242 &1.88&0.98&G& &73& & \\
   B0058-229 &1.24&0.95&Q&0.706&63&452&3E27\\
  B0100-277 &3.01&0.96&G& &67& & \\
  B0101-275 &1.86&1.06&G& &1.9& & \\
\\
  B0102-256 &1.95&1.07&G& &0.9& & \\
  B0103-243 &1.15&0.96&G& &20.8& & \\
   B0106-233 &1.13&0.85&Q&0.818&2.5&19&3E27\\
  B0106-291 &3.41&0.92&G& &1.9& & \\
  B0110-224 &1.31&1.05&G& &3.4& & \\
\\
   B0111-256 &0.98&0.68&Q&1.05&2.2&18&5E27\\
  B0112-209 &2.2&1.14&G& &38.1& & \\
  B0112-219 &0.98& &U& & & & \\
  B0113-245 &1.38&0.83&U& &50& & \\
  B0113-285 &1.69&0.77&G& &9.5& & \\
\\
  B0114-211 &10.64&0.87&G&1.41&1.9&16&1E29\\
  B0115-261 &2.58&0.74&G&0.268&10&41&5E26\\
  B0121-295 &1.46&0.96&G& &30.1& & \\
  B0122-255 &3.76&0.93&U& &48.1& & \\
   B0123-226 &1.54&0.58&Q&0.717&5.1&37&3E27\\
\\
  B0125-201 &1.08&0.89&G& &1.9& & \\
  B0125-216 &1.29&0.81&G&0.34&25.5&123&5E26\\
  B0127-276 &1.02&0.77&G&0.318&1.9&9&3E26\\
  B0128-264 &5.36&1.2&G& &33.4& & \\
   B0133-266 &1.19&0.97&Q&1.53&53.5&458&2E28\\
\\
  B0133-277 &0.97&1.03&G& &40.3& & \\
   B0136-231 &1.3&0.54&Q&1.895&12.8&109&2E28\\
  B0137-263 &1.46&0.83&G&0.16&77&210&1E26\\
  B0138-218 &0.96&0.97&G& &20.6& & \\
  B0139-273 &5.04&1.02&G&1.44&11.7&100&7E28\\
\\
  B0140-257 &1&1.24&G&2.64&3.4&28&8E28\\
  B0143-246 &1.51&0.86&G&0.716&52.5&379&3E27\\
  B0144-227 &1.18&0.81&G&0.6&1.9&13&2E27\\
  B0146-224 &1.65&0.81&G&0.36&14.3&72&7E26\\
  B0147-288 &1.28&0.95&G& &3.3& & \\
\\
  B0148-297 &7.04&0.82&G&0.41&138&749&4E27\\
  B0149-260 &1.05&0.95&G&0.144&96.5&241&6E25\\
  B0149-299 &2.42&0.83&G&0.603&17.1&115&3E27\\
  B0150-275 &1.94&0.91&G& &4& & \\
  B0152-209 &1.55&1.11&G&1.89&1&9&4E28\\
\\
  B0152-260 &1.31&0.87&G& &37& & \\
  B0155-212 &2.39&1.04&G&0.159&86&234&2E26\\
  B0155-225 &1.4&0.95&G& &37.6& & \\
  B0156-252 &1.39&1.04&G&2.09&6.8&57&5E28\\
  B0156-278 &0.96&0.76&G&0.33&5&24&3E26\\
\\
  B0201-214 &1.28&0.85&G&0.915&1.9&15&5E27\\
  B0203-209 &1.1&1.09&G&1.257&12&101&1E28\\
  B0205-223 &0.96&0.99&G& &24.2& & \\
  B0205-229 &1.77&0.83&G&0.68&34.4&243&3E27\\
  B0208-240 &1.87&0.74&G&0.23&67&244&3E26\\
\\
   B0209-237 &1.5&0.92&Q&0.68&18&127&3E27\\
   B0222-224 &2.36&0.95&Q&0.23&2.4&9&4E26\\
   B0222-234 &5.44&0.79&Q&1.617&15.5&133&8E28\\
   B0246-231 &2.44&0.7&Q&2.904&0.99&8&1E29\\
  B0209-282 &1.56&0.81&G&0.6&13.3&89&2E27\\
\\
  B0211-256 &1.07&1.05&G&1.3&2.4&20&1E28\\
  B0211-258 &1.19&0.82&G& &46& & \\
  B0216-250 &4.14&0.98&G& &55.2& & \\
  B0221-285 &3.86&1.08&G& &14.5& & \\
  B0223-245 &1.03&0.75&G&0.634&0.9&6&2E27\\
\\
  B0225-241 &2.16&0.78&G&0.52&5.2&32&2E27\\
  B0226-284 &1.063&0.69&G&0.21&58&197&1E26\\
  B0226-292 &1.12&0.97&G& &1.9& & \\
  B0230-245 &2.1&0.93&G&0.88&11.3&88&8E27\\
  B0231-235 &4.09&1.02&G&0.81&38.4&290&1E28\\
\\
  B0233-290 &1.96&1&G&0.725&4.9&36&5E27\\
  B0237-201 &1.5&0.85&G&1.03&19.1&155&8E27\\
  B0242-221 &0.98&1.12&G& &1.9& & \\
  B0245-263 &0.97&0.76&G&0.35&6.2&30&4E26\\
  B0245-297 &1.16&0.75&G&0.36&57&285&5E26\\
\\
  B0246-202 &1.69&0.8&G&0.58&1.9&12&2E27\\
  B0247-205 &1.12&1.53&G&0.32&2.9&13&4E26\\
  B0247-207 &3.3&0.97&G&0.085&200&315&6E25\\
  B0251-273 &0.98&1.06&G&3.16&3.9&30&1E29\\
  B0252-246 &1.38&1.07&G&1.3&33.6&284&1E28\\
\\
  B0253-206 &3.19&0.86&G&0.69&96&683&6E27\\
  B0253-259 &1.55&0.99&G& &1.9& & \\
  B0254-236 &5.87&1.1&G&0.509&33.4&205&6E27\\
  B0254-263 &1.93&1&G&0.31&6.5&29&6E26\\
  B0254-274 &1.01&1.19&G&0.48&37.4&223&9E26\\
\\
  B0255-247 &0.97&1.09&U& &1.5& & \\
  B0255-262 &1.33&1.12&G&0.36&8.4&42&6E26\\
  B0256-236 &2.19&0.9&U& &27.2& & \\
  B0259-252&1.14&0.98&U& &22& & \\
  B0259-252&1.11&0.83&U& &11& & \\
\\
  B0305-226 &4.95&0.88&G&0.268&88&359&1E27\\
  B0305-246 &1.21&1.07&G&1.265&6.4&54&1E28\\
  B0309-260 &0.95&1.41&G& &15& & \\
  B0312-271 &1.42&1.09&U& &1.9& & \\
  B0313-271 &1.82&1.14&G&0.216&227&789&2E26\\
\\
  B0315-205 &1.23&1.23&G& &21.1& & \\
   B0315-282 &1&0.52&Q&1.17&2&17&5E27\\
  B0316-257 &1.54&1.12&G&3.13&6.7&52&2E29\\
  B0319-298 &3.72&0.6&G&0.583&1.9&13&4E27\\
  B0320-263 &1.52&0.79&G& &22.5& & \\
\\
  B0320-267 &1.15&0.68&G&0.9&4.3&34&4E27\\
  B0324-228 &1.98&1.19&G&1.89&9.6&82&6E28\\
  B0325-260 &1.04&0.74&G&0.638&59&405&2E27\\
  B0326-288 &4.03&0.83&G&0.108&17&33&1E26\\
  B0327-261 &1.57&1.3&U& &60& & \\
\\
   B0328-272 &1.06&0.87&Q&1.803&18.1&155&2E28\\
  B0337-216 &1.51&0.84&G&0.414&1.1&6&9E26\\
  B0344-291 &2.14&0.72&G&0.137&4.9&12&1E26\\
  B0345-206 &1.13&0.98&G& &22.5& & \\
  B0346-297 &1.72&0.88&G&0.413&122&665&1E27\\
\\
  B0346-298 &1.01&0.96&G& &84& & \\
  B0349-211 &1.14&0.92&G&2.31&7.2&60&4E28\\
  B0349-278 &13.7&0.73&G&0.066&350&438&1E26\\
  B0350-279 &1.24&1.15&G&1.9&1.2&10&4E28\\
  B0353-207 &1.02&0.78&U& &1.9& & \\
\\
  B0354-263 &1.2&0.85&G& &12.6& & \\
  B0357-247 &2.16&0.87&G&0.205&30.3&101&3E26\\
  B0357-264 &1.5&0.54&U& &1.9& & \\
  B0400-247 &1.17&0.97&G&1.105&45.1&371&8E27\\
  B0406-244 &2.92&1.35&G&2.44&7.3&60&2E29\\
\\
   B0407-226 &1.24&1.01&Q&1.48&23.1&197&2E28\\
  B0412-204 &2.51&1&G&0.69&12.7&90&5E27\\
   B0413-210 &7.3&0.73&Q&1.63&5&43&1E29\\
   B0413-296 &3.71&1.07&Q&0.807&39&294&1E28\\
  B0415-221 &1.4&0.77&G& &14.6& & \\
\\
   B0418-288 &1.16&0.96&Q&0.85&1.99&15&4E27\\
  B0420-263 &2.82&0.7&G&0.131&219&506&7E29\\
   B0421-225 &1.72&0.78&Q&0.364&8.1&41&7E26\\
  B0422-249 &1.41&0.77&G& &4.9& & \\
  B0424-268 &3.25&0.87&G&0.47&22.5&132&3E27\\
\\
  B0428-236 &1.12&0.76&U& &8.9& & \\
  B0428-271 &1.72&0.83&G&0.84&46.4&355&5E27\\
  B0428-281 &2.49&0.89&G&0.65&61.9&429&4E27\\
  B0429-267 &1.32&1.11&G&1.27&6.9&58&1E28\\
  B0430-235&0.96&0.86&G&0.82&91&691&3E27\\
\\
   B0430-278 &0.95&0.76&Q&1.63&1.99&17&1E28\\
  B0431-250 &1.01&1.25&U& &16.4& & \\
  B0431-292 &1.67&1.57&G&0.406&1.2&6&1E27\\
  B0436-203 &0.97&0.53&U& &3.5& & \\
  B0436-294 &1.2&0.99&G&0.808&13.5&102&4E27\\
\\
   B0437-244 &1.28&0.93&Q&0.84&126&964&4E27\\
  B0437-253 &1.26&0.75&G& &20.2& & \\
  B0442-282 &18.85&0.97&G&0.147&85.6&218&1E27\\
  B0442-285 &1.33&1.16&G& &1.4& & \\
  B0442-289 &3.27&1.11&U& &11.9& & \\
\\
  B0445-221 &4.64&0.89&G& &1.9& & \\
   B0447-230 &0.96&0.83&Q&2.14&1.99&17&3E28\\
   B0450-221 &3.23&1.11&Q&0.898&14.3&112&1E28\\
  B0450-288 &1.44&0.94&G& &13.2& & \\
   B0454-220 &4.93&0.78&Q&0.533&84&529&5E27\\
\\
  B0457-235 &1.12&1.13&G&1.96&16.6&141&4E28\\
  B0457-247 &1.25&0.7&G&0.186&60&185&1E26\\
  B0458-208 &0.95&1.1&G& &33.8& & \\
  B0508-220 &5.1&0.81&G&0.16&38.5&105&3E26\\
  B0516-275 &1.37&0.81&G& &1.9& & \\
\\
  B0519-208 &7.34&1.18&G& &1.6& & \\
   B0522-215 &1.75&1&Q&1.83&2.5&21&4E28\\
  B0522-239 &1.08&0.8&G&0.5&23.4&142&9E26\\
  B0522-263 &1.36&0.91&G&0.29&4.9&21&3E26\\
  B0524-234 &1.26&0.87&G& &5.7& & \\
\\
  B0527-255 &1.54&0.88&G& &2& & \\
  B0528-212 &2.39&1.08&G& &34.5& & \\
  B0529-210 &1.33&0.96&G&0.42&40&220&8E26\\
  B0541-243 &3.61&0.98&G&0.523&20&125&4E27\\
  B0541-288 &0.99&0.79&G& &1.9& & \\
\\
  B0543-265 &2.63&0.88&G&0.85&23.9&184&9E27\\
   B0549-213 &1.7&0.83&Q&2.245&3.6&30&5E28\\
  B0551-226 &1&0.71&G& &70.5& & \\
  B0552-249 &1.31&0.86&G& &4.3& & \\
  B0555-229 &0.95&0.89&G& &10.2& & \\
\\
  B0556-281 &2.26&1.04&G& &6.8& & \\
  B0556-289 &2.39&0.59&G& &3.5& & \\
  B0557-235 &1.14&0.96&G& &21.2& & \\
  B0600-219 &1.04&1.09&G&1.71&4.2&36&2E28\\
  B0602-289 &1.37&1.03&G&0.56&37&239&2E27\\
\\
  B0614-295 &1.04&0.81&G& &9.1& & \\
  B0930-200 &3.26&1&G&0.769&20.2&150&9E27\\
  B0937-250 &1.37&1.07&G& &66.3& & \\
  B0938-205 &1.35&0.88&G&0.371&69.4&354&6E26\\
   B0941-200 &1.03&0.88&Q&0.715&47.7&344&2E27\\
\\
  B0943-242 &1.05&1.23&G&2.93&3.5&28&1E29\\
  B0946-237 &1.19&0.77&G& &7.8& & \\
  B0946-262 &1.74&1.02&G& &40.8& & \\
  B0947-217 &1&1.01&U& &24.2& & \\
  B0947-249 &4.98&1.07&G&0.854&69.1&532&2E28\\
\\
  B0949-206 &1.91&0.95&G&1.158&6&50&1E28\\
  B0950-239 &1.38&0.75&G& &19& & \\
  B0952-224 &1.71&0.62&G&0.228&6.2&22&2E26\\
  B0955-283 &1.43&0.91&G& &77.7& & \\
  B0955-288 &3.7&1&G&1.406&4.9&42&5E28\\
\\
  B0956-256 &2.21&0.94&G& &1.9& & \\
  B0958-227 &1.02&0.58&G&0.7&1.9&14&2E27\\
  B0959-225 &1.04&0.68&G&0.895&11.3&88&3E27\\
  B0959-236 &1.15&0.54&G& &41.4& & \\
  B0959-263 &1.72&0.82&G&0.677&39.3&277&3E27\\
\\
  B1002-215 &6.71&1.46&G&0.59&29.1&193&1E28\\
  B1002-216 &2.11&0.65&G&0.49&1.9&11&2E27\\
  B1006-214 &1.35&0.78&G&0.246&191&732&2E26\\
  B1006-286 &3.54&0.76&G&0.582&10.2&67&4E27\\
   B1006-299 &1.44&0.94&Q&1.064&18.3&149&8E27\\
\\
  B1008-233 &1.56&0.62&G&1.18&1.9&16&9E27\\
  B1009-259 &1.75&0.76&G& &15& & \\
   B1010-271 &1.42&1.09&Q&0.436&44.3&250&1E27\\
   B1011-282 &2.6&1.04&Q&0.255&64&252&5E26\\
  B1012-237 &2.09&1.02&G&0.993&33.1&266&1E28\\
\\
  B1014-200 &1.52&0.71&G& &1.9& & \\
  B1014-278 &1.5&0.92&U& &3.4& & \\
  B1017-220 &1.04&0.58&G&1.768&1.9&16&1E28\\
   B1019-227 &0.96&0.98&Q&1.55&2.2&19&1E28\\
  B1021-217 &1.18&1.06&G& &4& & \\
\\
  B1022-241 &1.04&0.8&G& &27.6& & \\
  B1022-250 &0.99&0.84&G&0.34&51.6&249&4E26\\
  B1022-299 &1.12&0.84&G&0.911&2&16&4E27\\
  B1023-226 &1.08&0.94&G&0.586&58.6&387&1E27\\
  B1023-243 &1.21&0.92&G& &6.1& & \\
\\
   B1025-229 &1.05&1.17&Q&0.309&188&849&3E26\\
   B1025-264 &1.53&0.63&Q&2.665&10.2&82&6E28\\
  B1025-270 &1.82&0.83&G&0.72&12.8&93&4E27\\
  B1025-293 &1.46&0.96&G& &22& & \\
  B1026-202 &1.95&0.88&G&0.566&62.1&403&2E27\\
\\
  B1027-225 &1.11&1.1&G&0.15&47&122&7E25\\
  B1029-233 &1.08&0.86&G&0.611&72&485&2E27\\
  B1033-251 &1.53&0.83&G&0.44&84.3&478&1E27\\
  B1033-259 &1.05&0.66&G& &0.9& & \\
  B1034-265 &1.37&0.92&U& &3.8& & \\
\\
  B1035-288 &1.77&0.84&G&1.276&7.9&67&1E28\\
  B1036-215 &0.98&0.82&G&0.585&41&271&1E27\\
  B1040-285 &1.09&0.99&G&1.63&6.4&55&2E28\\
  B1043-216 &1.83&1.13&G&1.105&3.1&26&1E28\\
  B1048-211 &1.5&1.07&U& &1.9& & \\
\\
  B1048-238 &1.31&0.76&G&0.206&82&275&2E26\\
  B1048-272 &2.41&0.79&G&1.558&1.9&16&3E28\\
  B1049-201 &4.46&0.98&G&1.116&4.4&36&3E28\\
  B1051-274 &0.97&1.13&G& &38.8& & \\
   B1052-272 &2.05&1.07&Q&1.103&76.5&630&1E28\\
\\
   B1055-242 &1.95&0.51&Q&1.09& & &9E27\\
  B1056-272 &0.97&0.97&G&0.25&7.1&28&2E26\\
  B1103-208 &7.64&0.99&G&1.12&9.7&80&5E28\\
   B1106-227 &1.81&0.73&Q&1.875&0.99&8&3E28\\
  B1106-258 &0.97&1.26&G&2.43&3.6&30&6E28\\
\\
  B1107-218 &1.04&0.95&G& &53.8& & \\
  B1107-227 &2.89&1.12&G& &68.5& & \\
  B1107-272 &1&0.97&U& &6.8& & \\
  B1108-212 &1.01&0.84&G& &13.9& & \\
  B1110-217 &2.69&0.54&G& &1.9& & \\
\\
  B1112-239 &1.46&1.06&G&1.538&30.7&263&2E28\\
  B1114-217 &1.32&0.94&G& &13.6& & \\
   B1114-220 &1.61&0.72&Q&2.282&0.99&8&5E28\\
  B1117-217 &1&1.08&G& &36.4& & \\
   B1117-248 &2.69&0.54&Q&0.462& & &2E27\\
\\
   B1121-238 &1.57&1.09&Q&0.675&46&324&3E27\\
  B1126-246 &1.1&0.74&G&0.155&51.7&138&7E25\\
  B1126-258 &1.13&1.01&G&0.979&8.1&65&6E27\\
  B1126-290 &2.42&0.74&G&0.41&105&570&1E27\\
  B1128-268 &1.09&1.05&G&1.43&7&60&1E28\\
\\
  B1129-250 &0.98&0.91&G&1.065&32&261&6E27\\
  B1131-269 &2.36&1.32&G&1.711&0.9&8&7E28\\
  B1132-258 &2.56&0.88&U& &0.9& & \\
  B1136-211 &1.1&1.02&G&0.87&25.5&197&4E27\\
  B1137-257 &0.95&0.62&G& &2.9& & \\
\\
  B1138-262 &4.12&1.34&G&2.17&11.1&93&2E29\\
  B1139-285 &6.81&0.85&G&0.85&13&100&2E28\\
  B1142-206 &1.42&1.03&G& &54.5& & \\
  B1142-242 &1.15&0.9&G& &13.6& & \\
  B1145-248 &1.04&0.76&G& &17.4& & \\
\\
   B1151-298 &1.44&0.85&Q&1.376&4&34&1E28\\
  B1152-204 &0.98&0.81&G& &13.7& & \\
  B1153-231 &1.94&0.94&G& &45& & \\
  B1155-214 &1.27&0.95&G& &32& & \\
   B1156-221 &2.66&0.61&Q&0.563& & &3E27\\
\\
  B1158-275 &0.98&0.9&G& &9.5& & \\
   B1202-262 &3.55&0.53&Q&0.786&15&112&8E27\\
   B1208-277 &1.58&0.83&Q&0.828&43.4&331&5E27\\
  B1210-290 &1.08&0.89&G& &6& & \\
  B1211-259 &1.1&0.83&G& &8.4& & \\
\\
  B1211-272 &0.97&0.98&G&1.893&2.6&22&2E28\\
  B1212-204 &1.32&0.94&G& &42& & \\
   B1212-275 &0.96&0.96&Q&1.656&3.5&30&2E28\\
   B1217-209 &1&0.94&Q&0.814&29.8&226&3E27\\
  B1217-276 &1.03&1.19&G&1.899&7.5&64&3E28\\
\\
  B1219-264 &1.13&1.22&G& &93& & \\
   B1222-293 &1.33&0.76&Q&0.816&29.4&223&4E27\\
  B1224-208 &1&1.09&G& &62& & \\
   B1224-262 &3.18&0.83&Q&0.768&1.99&15&8E27\\
  B1226-211 &3.28&0.8&G&0.191&29.3&92&3E26\\
\\
   B1226-297 &1.2&0.84&Q&0.749&64.5&474&3E27\\
  B1230-244 &1.93&0.93&G&0.257&3.4&13&4E26\\
   B1232-249 &5.1&0.83&Q&0.352&109&537&2E27\\
  B1235-226 &1.92&0.89&G&0.778&7.1&53&5E27\\
  B1236-200 &1.04&0.96&G& &3& & \\
\\
  B1238-236 &1.11&0.93&G&0.9&5.9&46&4E27\\
  B1239-256 &0.96&1.01&G& &12.8& & \\
  B1240-209 &4.58&0.87&G&0.42&18.8&104&3E27\\
  B1240-271 &1.45&1.01&G& &1.9& & \\
  B1241-275 &1.67&1.33&G& &17& & \\
\\
  B1241-291 &1.37&1.1&G& &17.6& & \\
  B1245-292 &1.81&1&G& &21& & \\
  B1246-206 &1.18&0.78&G& &46.2& & \\
  B1246-231 &1.17&0.63&G&0.68&1.9&13&2E27\\
   B1247-290 &1.87&0.91&Q&0.77&57.6&428&5E27\\
\\
  B1254-268 &1.14&0.74&G&0.135&24.5&58&5E25\\
  B1255-282 &1.15&0.93&G& &39.1& & \\
   B1257-230 &3.23&1.05&Q&1.109&52&428&2E28\\
  B1258-211 &1.52&1.19&G&1.58&2.6&22&3E28\\
  B1259-200 &3.57&1.15&G&1.58&9.9&85&7E28\\
\\
   B1301-251 &1.16&0.79&Q&0.952&8.7&69&5E27\\
  B1302-206 &1.2&1.06&G& &18.2& & \\
  B1303-215 &1.46&0.76&G&0.12&70&150&5E25\\
   B1303-250 &1.51&0.8&Q&0.738&38.5&281&3E27\\
  B1306-262 &1.19&0.86&G& &56& & \\
\\
  B1307-217 &1.09&0.92&G& &14& & \\
  B1308-220 &22.21&1.19&G&0.8&1.1&8&7E28\\
  B1309-201 &1.79&0.82&G& &1.9& & \\
  B1309-211 &1.67&0.83&G&0.3&53.7&238&5E26\\
   B1309-216 &1.03&0.69&Q&1.49&3&26&1E28\\
\\
  B1309-294 &1.22&0.82&G&0.67&3.2&22&2E27\\
   B1311-270 &1.77&0.82&Q&2.186&19.5&164&5E28\\
  B1312-274 &1.78&1.17&G& &32.4& & \\
  B1313-248 &2.14&0.9&G&0.74&27.7&203&5E27\\
  B1313-267 &1.05&1.01&U& &13.5& & \\
\\
  B1324-262 &1.39&1.1&G&2.28&1.6&13&6E28\\
  B1325-222 &2.12&0.99&G&0.4&3.8&20&1E27\\
  B1325-257 &1.16&0.93&G&0.62&45.4&308&2E27\\
   B1327-214 &5.63&0.82&Q&0.528&31&194&6E27\\
  B1328-257 &4.8&1.03&G& &5& & \\
\\
  B1329-257 &3.73&0.89&G&0.19&48.8&153&4E26\\
  B1331-214 &1.05&0.97&G& &22& & \\
  B1336-276 &1.03&1.05&G& &3.5& & \\
  B1344-216 &2.7&0.87&G&0.33&1.9&9&9E26\\
  B1346-252 &1.27&1.26&G&0.125&55&122&5E25\\
\\
   B1349-265 &3.59&0.63&Q&0.934&1.99&16&1E28\\
   B1351-211 &2.29&0.94&Q&1.262&11&93&2E28\\
  B1351-235 &1.77&1.11&U& &10.6& & \\
  B1353-216 &1.44&1.13&G& &23& & \\
  B1353-245 &1.38&0.96&G& &3.4& & \\
\\
   B1355-215 &1.87&0.8&Q&0.832&4.2&32&6E27\\
   B1355-236 &1.44&0.95&Q&1.604&16&137&2E28\\
  B1357-217 &0.96&0.96&G& &15.2& & \\
  B1358-214 &1.48&1.08&G&0.5&90&548&1E27\\
   B1359-281 &2.3&0.52&Q&0.802&1.4&11&5E27\\
\\
  B1401-296 &3.28&0.83&G& &12.7& & \\
  B1402-253 &2.37&0.98&G&0.74&5&37&6E27\\
   B2021-208 &1.65&1.16&Q&1.2&24.5&205&2E28\\
   B2024-217 &2.45&0.88&Q&0.459&31&180&2E27\\
   B2025-206 &1.94&1.12&Q&1.4&32.5&277&3E28\\
\\
  B2025-218 &1.28&1.07&G&2.63&4&32&8E28\\
  B2028-223 &2.58& &U& & & & \\
  B2028-293 &1.38&0.99&G&0.498&4.9&30&1E27\\
   B2030-230 &6.45&1&Q&0.132&70&163&3E26\\
   B2035-203 &1.87&0.77&Q&0.516&64&396&2E27\\
\\
  B2036-254 &1.19&1.06&G&2&5.9&50&4E28\\
   B2037-234 &0.96&0.93&Q&1.15&15&124&7E27\\
  B2038-280 &1.47&1.23&G&0.39&150&790&8E26\\
  B2039-236 &1.63&0.86&G&0.621&45&305&2E27\\
  B2039-291 &3.02&0.91&G& &15.1& & \\
\\
  B2040-219 &1.2&1.08&G&0.204&7.1&24&1E26\\
   B2040-236 &1.05&0.61&Q&0.704&56&401&2E27\\
  B2042-293 &1.18&0.92&G& &65& & \\
  B2044-272 &1.12&0.89&U& &0.9& & \\
  B2045-245 &1.99&1&G&0.73&77&560&5E27\\
\\
  B2045-256 &3.08&1.21&G& &0.9& & \\
  B2045-260 &0.95&1.12&G& &25& & \\
  B2048-272 &1.98&1.27&G&2.06&5.3&45&9E28\\
  B2052-253 &1.09&1.23&G&2.6&18.1&147&8E28\\
  B2053-201 &6.37&0.8&G&0.155&29.7&79&4E26\\
\\
  B2057-286 &2.34&0.93&G&0.605&12.3&82&3E27\\
  B2058-237 &1.47&0.96&G& &1.9& & \\
  B2059-228 &1.41&0.93&G& &28.4& & \\
  B2100-280 &2.37&0.79&G& &12.5& & \\
  B2101-214 &0.96&0.73&G&0.198&31.9&104&1E26\\
\\
  B2104-242 &1.8&1.35&G&2.49&21.8&179&1E29\\
  B2104-256 &28.1&0.79&G&0.037&270&196&9E25\\
  B2104-290 &0.97&1&G& &21.6& & \\
  B2105-238 &1.47&1.03&G& &33& & \\
  B2107-285 &1.53&0.66&G& &43.8& & \\
\\
   B2111-259 &5.27&0.91&Q&0.602&9&60&8E27\\
  B2111-275 &0.95&1.38&G& &19& & \\
  B2113-211 &9.05&0.96&G&0.698&40.4&289&2E28\\
  B2115-253 &1.23&1.19&G&1.114&1.7&14&1E28\\
  B2116-250 &1.74&0.89&G&0.467&46&270&1E27\\
\\
  B2116-294 &1.01&0.88&U& &77& & \\
  B2117-269 &2.6&0.77&G&0.103&31.3&59&7E25\\
  B2118-266 &1.17&0.58&G&0.343&80&388&4E26\\
  B2118-296 &0.97&1.18&G& &8& & \\
   B2122-238 &1.05&0.71&Q&1.774&1.6&14&2E28\\
\\
  B2123-292 &2.12&0.99&G& &24.2& & \\
  B2125-237 &2.21&1.09&G&0.95&1.9&15&1E28\\
  B2126-230 &2.58&1.05&G& &35& & \\
   B2128-208 &6.15&0.97&Q&1.62&1.99&17&1E29\\
  B2131-241 &1.04&0.92&G& &1.9& & \\
\\
  B2132-236 &0.95&0.95&G&0.81&54.5&412&3E27\\
  B2135-209 &9.76&0.79&G&0.635&1.9&13&1E28\\
  B2135-257 &1.38&0.9&G&1.31&1.9&16&1E28\\
   B2136-251 &1.2&0.57&Q&0.94&1.99&16&4E27\\
  B2136-261 &2.93&1.01&G& &29.8& & \\
\\
  B2137-279 &1.21&0.94&G&0.64&59.2&407&2E27\\
  B2139-292 &1.66&1.09&G&2.55&6.8&56&1E29\\
  B2144-236 &1.34&1.12&G& &13.2& & \\
  B2144-279 &1.09&0.83&G& &37.8& & \\
  B2148-228 &1.42&0.99&G&0.85&21&161&5E27\\
\\
   B2149-200 &5.12&0.9&Q&0.424&2&11&3E27\\
  B2149-287 &5.68&0.64&G&0.479&1.9&11&4E27\\
  B2150-202 &2.68&1.2&G& &39.4& & \\
  B2151-283 &1.49&1.24&U& &10.4& & \\
  B2154-293 &1.01&0.86&G&0.63&1.9&13&2E27\\
\\
  B2155-255 &0.98&1.48&G& &2.7& & \\
   B2156-245 &1.39&0.87&Q&0.862&1.99&15&5E27\\
   B2158-206 &1.15&0.87&Q&2.272&0.99&8&4E28\\
  B2159-201 &1.74&1.27&G& &2.7& & \\
  B2200-251 &1.27&1&G& &11.3& & \\
\\
  B2201-272 &1.47&0.9&G&0.93&5.8&46&6E27\\
  B2204-202 &2.9&1.18&G&1.61&2&17&6E28\\
  B2206-237 &3.78&0.51&G&0.087&1.9&3&7E25\\
  B2206-251 &2.04&0.93&G&0.158&104&281&1E26\\
  B2210-283 &1.42&0.64&G& &1.9& & \\
\\
  B2211-252 &1.25&0.94&G& &17.7& & \\
   B2211-251 &2.3&0.98&Q&2.508&2.4&20&1E29\\
   B2213-283 &2.54&0.98&Q&0.946&58&460&1E28\\
  B2216-206 &1.23&1.03&G&1.148&88.6&735&9E27\\
  B2216-281 &6.24&1.03&G&0.657&1.9&13&1E28\\
\\
  B2217-251 &2.45&1.07&G& &1.9& & \\
  B2222-277 &1.36&0.82&G& &8.4& & \\
  B2224-273 &1.15&1.32&G&1.68&0.9&8&3E28\\
  B2226-224 &1.25&0.89&G&0.38&10.4&54&6E26\\
  B2226-297 &1.2&1&G&0.73&4.9&36&3E27\\
\\
   B2227-214 &1.81&1&Q&1.41&15.1&129&2E28\\
  B2229-228 &1.1&0.86&G&0.542&6.6&42&1E27\\
  B2230-206 &1.01&0.66&G&0.6&5.8&39&1E27\\
  B2232-232 &2.19&0.79&G&0.87&15&116&7E27\\
   B2232-272 &1.11&0.89&Q&1.495&16&137&1E28\\
\\
  B2236-264 &1.23&0.83&G&0.43&25.7&144&8E26\\
  B2238-216 &1.07&1&G&0.401&7.1&38&6E26\\
  B2247-232 &3.3&1.02&G&1.33&9.3&79&4E28\\
  B2247-248 &1.05&0.9&G&1.63&13.1&112&2E28\\
  B2248-223 &1.3&0.89&G&0.307&71&319&4E26\\
\\
  B2250-210 &1.11&0.85&G&0.72&2.9&21&2E27\\
  B2254-248 &1.81&0.68&G&0.54&25&159&2E27\\
  B2255-228 &1.12&1.35&G& &0.9& & \\
  B2256-207 &1.21&1.02&G&0.87&32.7&253&5E27\\
   B2256-217 &1.33&1.23&Q&1.779&33.2&284&4E28\\
\\
   B2257-270 &1.45&0.62&Q&1.476&0.99&8&1E28\\
  B2303-253 &2.52&1.07&G&0.73&18.6&135&6E27\\
  B2304-257 &1.27&0.94&G& &1.2& & \\
  B2307-282 &3.1&1&G& &37.6& & \\
  B2308-214 &0.99&0.83&G&0.151&58.5&152&6E25\\
\\
  B2311-222 &2.24&0.78&G&0.434&80&450&1E27\\
  B2313-277 &1.9&0.97&G&0.614&44.2&298&3E27\\
  B2314-211 &1.15&1.08&G& &17& & \\
  B2317-223 &1.793&1.24&G& &34.9& & \\
  B2317-277 &5.44&0.73&G&0.173&210&612&4E26\\
\\
  B2318-244 &2.42&0.93&G&1.12&25&206&2E28\\
  B2320-269 &0.97&0.9&G&0.99&1.9&15&5E27\\
  B2322-275 &3.07&0.72&G&1.27&22.3&188&2E28\\
  B2324-259 &1.44&0.72&G&0.286&21.7&93&3E26\\
  B2325-213 &3.07&0.94&G&0.58&79&519&4E27\\
\\
  B2326-254 &1.91&0.98&G& &2.4& & \\
  B2327-215 &2.05&0.88&G&0.28&4.9&21&5E26\\
  B2329-251 &2.75&1.12&G& &18.4& & \\
   B2338-233 &1&0.82&Q&0.715&29.5&213&2E27\\
   B2338-290 &1.28&0.75&Q&0.446&73&417&8E26\\
\\
  B2340-219 &2.35&1.06&G&0.766&27.1&201&7E27\\
  B2341-244 &1.55&0.74&G&0.59&2.1&14&2E27\\
  B2343-243 &1.85&0.8&G&0.6&48.3&323&2E27\\
  B2348-235 &1.47&0.83&G&0.952&68.7&546&6E27\\
   B2348-252 &4.41&1.13&Q&1.386&33.5&285&6E28\\
\\
  B2351-222 &1.41&1&G& &24.1& & \\
  B2351-234 &1.61&0.92&G&1.03&28.8&234&9E27\\
  B2355-214 &2.09&1.23&G&1.41&2.8&24&3E28\\
  B2359-259 &1.02&0.81&G& &1.9& & \\
\end{longtable}
\end{document}